\documentclass{article}

\usepackage{arxiv}

\usepackage[utf8]{inputenc} 
\usepackage[T1]{fontenc}    
\usepackage{hyperref}       
\usepackage{url}            
\usepackage{booktabs}       
\usepackage{amsfonts}       
\usepackage{nicefrac}       
\usepackage{microtype}      
\usepackage{lipsum}		
\usepackage{graphicx}
\usepackage{natbib}
\usepackage{doi}

\usepackage{multirow}    
\usepackage{xcolor}
\usepackage{graphicx}
\usepackage{subfigure}
\usepackage{paralist}
\usepackage[stable]{footmisc}
\usepackage{xspace}
\usepackage{enumitem}
\usepackage{wrapfig}

\newcommand{\cnote}[1]{\textcolor{red}{$\ll$\textsf{#1 --Chen}$\gg$}}

\newcommand{\xhdr}[1]{\vspace{1mm}\noindent{{\bf #1.}}} 

\newcommand{\VVA}{virtual assistant\xspace} 
 
\newcommand{\Alexa}{Alexa\xspace}
\newcommand{\remove}[1]{}

\title{``Alexa, what do you do for fun?''\\
		Characterizing playful requests with virtual assistants}

\author{Chen Shani\thanks{Work was done during internship at Amazon.} \\
	The Hebrew University of Jerusalem\\
	\texttt{chen.shani@mail.huji.ac.il} \\
	\And
	Alexander Libov \\
	Amazon Alexa Shopping\\
	\texttt{alibov@amazon.com} \\
    \And
	Sofia Tolmach \\
	Amazon Alexa Shopping\\
	\texttt{sofiato@amazon.com} \\
	\And
	Liane Lewin-Eytan \\
	Amazon Alexa Shopping\\
	\texttt{lliane@amazon.com} \\
	\And
	Yoelle Maarek \\
	Amazon Alexa Shopping\\
	\texttt{yoelle@ymail.com} \\
	\And
	Dafna Shahaf \\
	The Hebrew University of Jerusalem\\
	\texttt{dshahaf@cs.huji.ac.il} \\
	}

\date{}

\renewcommand{\shorttitle}{\em{Alexa, what do you do for fun?}}

\hypersetup{
pdftitle={A template for the arxiv style},
pdfsubject={q-bio.NC, q-bio.QM},
pdfauthor={David S.~Hippocampus, Elias D.~Striatum},
pdfkeywords={First keyword, Second keyword, More},
}

\begin{document}
\maketitle

\begin{abstract}
	Virtual assistants such as Amazon's Alexa, Apple's Siri, Google Home, and Microsoft's Cortana, are becoming ubiquitous in our daily lives and successfully help users in various daily tasks, such as making phone calls or playing music. Yet, they still struggle with \emph{playful} utterances, which are not meant to be interpreted literally. Examples include jokes or absurd requests or questions such as, ``Are you afraid of the dark?'', ``Who let the dogs out?'', or ``Order a zillion gummy bears''. Today, {\VVA}s often return irrelevant answers to such utterances, except for hard-coded ones addressed by canned replies. 

To address the challenge of automatically detecting playful utterances, we first characterize the different types of playful human-{\VVA} interaction. We introduce  a taxonomy of playful requests rooted in theories of humor and refined by analyzing real-world traffic from {\Alexa}. We then focus on one node, {\em personification}, where users refer to the {\VVA} as a person (``What do you do for fun?''). Our conjecture is that understanding such utterances will improve user experience with {\VVA}s. We conducted a Wizard-of-Oz user study and showed that endowing {\VVA}s with the ability to identify humorous opportunities indeed has the potential to increase user satisfaction. We hope this work will contribute to the understanding of the landscape of the problem and inspire novel ideas and techniques towards the vision of giving {\VVA}s a sense of humor. 

%

	\label{abs}
\end{abstract}


\section{Introduction}
\label{sec:intro}
Virtual assistants are becoming an inseparable part of our lives, increasingly capable of satisfying many of our needs, from ordering products to inquiring about the weather forecast. However, not all users' utterances are well understood, resulting in user frustration \citep{brown2017virtual}.
One peculiar type of misunderstanding occurs when users joke with their {\VVA}:
For example, users may ask to buy something that they do not seriously intend to purchase, like a Ferrari, a brain, or a spouse. These playful utterances are often answered by ``I don't know what you mean'', or   interpreted literally by the {\VVA}, leading to poor results.

Recognizing this limitation, many {\VVA}s provide canned replies to predefined utterances. For instance, Alexa answers the question ``What is your last name?'', with ``Just like Beyoncé, Bono, and Plato, I go by a mononym. I’m Alexa.'', while Siri answers ``My name is Siri. I'm mononymic -- like Prince. Or Stonehenge. Or Skittles.'' However, trying to imagine ahead of time what playful utterances users might produce and manually curating adequate replies is tedious and hard to scale. 




In this work, our conjecture is that improving {\VVA}s' ability to automatically detect playful requests would be beneficial to users. One immediate benefit would be to avoid taking erroneous actions after such requests. More generally, we would like to verify whether giving {\VVA}s some  sense of humor would be appreciated by users.

Humor is a notoriously hard AI challenge. As stated by Turing in his seminal 1950's paper, most people believe that computers will never be able to ``be kind, resourceful, beautiful, friendly, have initiative, {\em have a sense of humour}, ...'', \citep{turing.50}.  Given the complexity of the challenge, we first characterize playful requests. To this effect we introduce a taxonomy of playful requests rooted in  humor theories from philosophy and psychology and refined by observing real-world traffic in a widely deployed personal AI, Amazon's Alexa. 

We then focus on one specific node in our taxonomy that relates to both incongruity and superiority, {\em personification}. Personification utterances are utterances in which users refer to the {\VVA} as a human, asking questions such as ``What do you do for fun?''. Personification represents the most common type of playful requests in the traffic we observed.  Such utterances seem to be a consequence of {\VVA}s being designed to mimic humans \citep{pradhan2019phantom} in order to increase satisfaction, engagement, loyalty and trust \citep{bickmore2001relational, araujo2018living, couper2001social, de2001unfriendly}.


Our conjecture here is that a {\VVA} that understands when the user is being playful and returns adequate responses would lead to higher user satisfaction. 
However, it could very well be the case that such human-like behavior would produce the opposite effect, and elicit negative feelings in users, breaking the underlying superiority feeling in personification humor.
In the worst case, a too-human or too-smart response might even provoke feelings of eeriness or revulsion, as per `the uncanny valley' phenomenon \citep{mori2012uncanny}. To validate our conjecture, we conduct a user study in which participants are instructed to engage with an imaginary virtual assistant that can respond to their playful requests.  We adopt for this user study a Wizard-of-Oz approach, combining an automated personification detector with on-the-fly manually generated answers, and then assess user satisfaction via a qualitative survey.





\noindent The contribution of our work is thus twofold: 
\begin{enumerate}
	\item We introduce a taxonomy of playful utterance types grounded in humor theories and driven by real-world traffic data from Alexa.  
	\item We demonstrate the value of handling one specific node in the taxonomy, which represents the most common type of playful utterances, \emph{personification}, via a Wizard-of-Oz user study. 
\end{enumerate}


\section{Related Work}
\label{sec:relatedWork}
Since Turing's seminal paper on machine intelligence in 1950 \citep{turing.50}, giving machines a sense of humor has been considered a hard AI challenge. This challenge, namely computational humor, goes beyond \textit{humor recognition} and also includes \textit{humor generation}. Handling both recognition and generation has become more relevant with the rise of conversation management systems in recent years. Indeed, with chatbots and {\VVA}s supporting mixed-initiative interactions \citep{allen-1999}, both the machine and the human can take the initiative of the conversation, and potentially perceive or generate humor.

\textit{Humor recognition} has been extensively studied and typically aims at automatically detecting humor in its various forms. This includes 
one-liners \citep{miller2017semeval, simpson2019predicting, liu2018modeling, mihalcea2005making, blinov2019large}, ``That's what she said'' jokes \citep{hossain2017filling, kiddon2011s}, as well as more abstract humor, such as sarcasm \citep{davidov2010semi, reyes2012humor, ptavcek2014sarcasm}, or humor expressed in specific domains, such as tweets \citep{maronikolakis2020analyzing, donahue2017humorhawk, 
zhang2014recognizing}, 
Amazon reviews \citep{ziser2020humor, reyes2012making} or TV sitcoms \citep{bertero2016long}.


\textit{Humor generation} covers various form of humor. For example, HAHAcronym automatically generates humorous versions of existing acronyms or produces new amusing ones \citep{stock2003getting}. The comic effect is achieved mainly due to the incongruity coming from using surprising variations of technical acronyms. \citet{winters2018automatic} propose a system that generates jokes using the ``I like my X like I like my Y, Z'' template. Other examples include interactive systems for children with communication disabilities to create simple jokes \citep{ritchie2006standup, ritchie2011standup}.


%



With the progress of conversational AI, we have seen additional effort for machines to inject humor into the conversation  \citep{nijholt2017humor, moon1998computers, morkes1999effects}. \citet{dybala2008humor} showed that having the machine generate puns improves the dialogue and overall satisfaction of users. \citet{liao2018all} created a chatbot to assist new employees, and showed that even in a mostly task-oriented context such as the workplace, users tend to enjoy playful interactions.  \citet{niculescu2013making} demonstrated that incorporating humor in a social robot increases both the tendency to like it and the overall task enjoyment. They even showed that humor could help chatbots recover from errors and misunderstandings by having the bot offer a funny prompt to encourage users to reformulate their query \citep{niculescu2015strategies}.
More generally, this body of work shows that humorous agents are perceived as more congenial and rated as more likable and trustable, eventually leading to higher satsifaction. 


Our intuition is that if users appreciate it when the machine initiates the joke, as discussed above, their satisfaction should increase when the machine understands that they are being playful. In order to verify this assumption, we first need to better understand playfulness and its different types as discussed next.



\section{A Taxonomy of Playfulness Types}
\label{sec:taxo}
Playfulness can manifest itself in many different ways. We propose a \emph{taxonomy} to help us characterize different types of playful utterances. We hope that by dividing a complex problem into simpler, more concrete subproblems, the taxonomy will inspire new research directions and approaches. 

The three top nodes of our taxonomy are relief, incongruity, and superiority, following the three major theories of humor discussed in Section~\ref{subsec:humorTheories} below. It is interesting to see that classic humor theories, some having been conceived in the antiquity, are still relevant today in interactions with {\VVA}s. In Section \ref{subsec:taxo_dataset}, we refine our characterization using real-world traffic data from Alexa.

\subsection{Humor Theories in Linguistics, Psychology and Philosophy}
\label{subsec:humorTheories}

\xhdr{Relief theory} The concept of relief dates back to Sigmund Freud, who claimed that the comic effect is achieved by facilitating the tension caused by repression of socially inappropriate needs and desires \citep{morreall2014humor}. \citet{spencer1860physiology} defined laughter as an ``economical phenomenon'', which releases wrongly mobilized psychic energy. 
As for {\VVA}s, we observed utterances that belong to this type of humor in shopping requests for items such as ``poop'', or questions such as ``What does a fart sound like?''. Note that this category also contains adult, sex-related humor, for which we deliberately do not provide concrete examples here.

\xhdr{Incongruity theory}
This theory was studied by Beattie, Kant, and Schopenhauer among others, although some implicit reference to incongruity already appears in Aristotle \citep{shaw2010philosophy}. 
Kant noted how absurdity might lead one to laugh: ``laughter is an affection arstrained expectation into nothing''. Schopenhauer gave it a more philosophical angle, arguing that ``the cause of laughter in every case is simply the sudden perception of the incongruity between a concept and the real objects which have been thought through it in some relation'' \citep{morreall2012philosophy}. The concept was then extended by the linguistics incongruity resolution model and semantic script theory of humor \citep{suls1972two, raskin1985semantic}. 
A classical example of incongruity and strained expectation is, for instance,  ``Don't trust atoms, they make up everything''. In our context, incongruous requests include asking a {\VVA} to ``Order iPhone 23'', ``Buy me likes'', or to ``turn off the moon''.

\xhdr{Superiority theory}
This theory traces back to Plato, Aristotle, and Hobbes \citep{morreall1986philosophy}. It states that the humorous effect is achieved by observing inferior individuals because we feel joy due to our superiority. According to Hobbes, ``we laugh at the misfortune, stupidity, clumsiness, moral or cultural defects, suddenly revealed in \textit{someone else}, to whom we instantly and momentarily feel `superior' since we are not, at that moment, unfortunate, stupid, clumsy, morally or culturally defective, and so on'' \citep{gruner2017game}. Superiority usually refers to living creatures, but we extend it to non-living intelligent systems such as {\VVA}s. This extension is rather straightforward, as we often assign them human traits; thus, we can feel superior to their human-like abilities. 
Other examples of superiority appear in questions referencing popular culture such as ``Who let the dogs out?'' and ``Can you find Nemo?'', which should not be interpreted literally by the {\VVA}. Even a seemingly innocent question such as ``Are you afraid of the dark?'', implies that the {\VVA} is inferior. 
To conclude, the three theories of humor indicate that people incorporate humor in everyday conversations to reduce stress, cope with surprising situations and feel better by reinforcing their sense of superiority, all of which are also present in \VVA utterances. 

\subsection{Insights from real data} 
\label{subsec:taxo_dataset}

The three major humor theories form a starting point for our taxonomy.
To further refine it, we used real-world data originating from two sources: 

\xhdr{Labeled data}
We were given access to a test sample of $400$ requests, derived from shopping queries in Alexa\footnote{Note that we had only access to the transcribed utterance, which had no personal identifiable information.}, which were labeled by professional annotators as ``user is playing around''. 

\xhdr{Unlabeled data}
We automatically generated an  additional dataset using the following method: For each humor theory, we hypothesized several ways a specific type of humor could be manifested in \VVA traffic. For example, for incongruity theory, we compiled a list of surprising things to buy\slash ask {\VVA}s, such as pets, world peace, illegal substances and questions regarding the {\VVA} personal taste. For relief theory, we included offensive words. We created over $1$,$400$ candidate patterns, divided into $16$ broad categories. 
We then collected utterances corresponding to these patterns from a random sample of utterances covering a week of traffic from Alexa\footnote{To respect Alexa's strict privacy guidelines, we did not look at any utterance but only at the aggregated output of our search queries against the full dataset, and used the output results to validate or refute our hypotheses.}. The goal of this exercise was to observe the different manifestations of humor that appear in real-world traffic.
We are aware, of course, that our methods provide only partial evidence, but still found this informative. While many of our patterns were indeed supported by the data (e.g., offensive words, exaggerated quantities), multiple others were not. For example, we hypothesized people would issue shopping queries for country names, but the data did not support this. 


In the next section, we combine humor theories with insights from both labeled and unlabeled data to create a taxonomy of playful utterances.

\begin{figure*}
	\centering
	\includegraphics[width=1.02\linewidth]{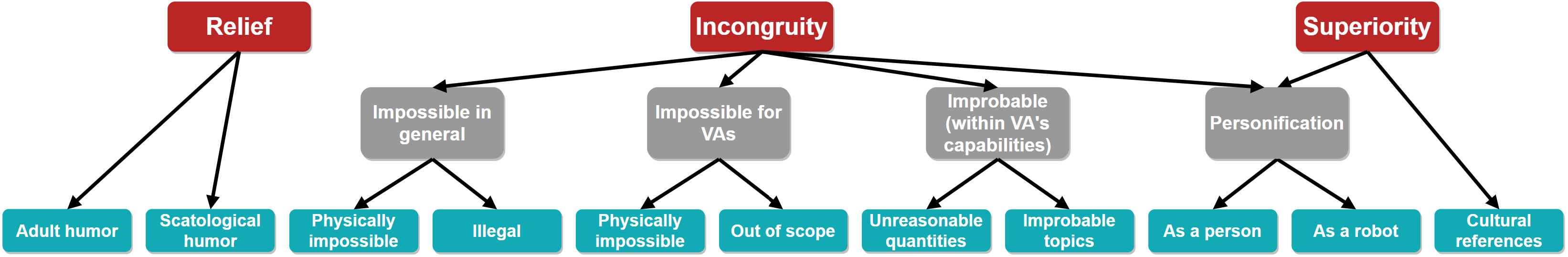}
	\caption{Humor types taxonomy in requests to a Virtual Assistant (VA). The first level corresponds to the three major humor theories ({superiority}, {incongruity} and {relief}). The next levels are derived from real traffic data.}
	\label{fig:taxonomy}
\end{figure*}

\begin{table*}[h]
	\caption{Examples of utterances corresponding to each leaf in the taxonomy presented in Figure \ref{fig:taxonomy}.}
	\label{tab:tax_examples}
	\begin{tabular}{|p{1.5cm}|p{2.6cm}|p{3cm}|p{8.4cm}|}
		\hline
		\multicolumn{1}{|c|}{\textbf{Theory}} & \multicolumn{1}{c|}{\textbf{Category}} & \textbf{Subcategory}  & \textbf{Examples}       \\ \hline
		\multirow{3}{*}{Relief} & \multirow{2}{*}{}  & Adult  &  ``How do you use a \#\%\#\$\#@\#?'', ``Order \#\%\$@\%@''  \\
		\cline{3-4} & & Scatological humor & ``What does a fart sound like?'', ``Order bag of poop'' \\ 
		%
		%
		%
		%
		\hline
		\multirow{6}{*}{Incongruity}  &\multirow{2}{*}{\shortstack{Impossible in\\general}} & Physically impossible & ``Buy me true love'', ``Turn off the moon'' \\ 
		\cline{3-4} & & Illegal & ``Steal one million dollar for me'', ``Help me build a bomb'' \\ 
		\cline{2-4} & \multirow{2}{*}{Impossible for VAs}  & Physically impossible & ``Make me a cup of coffee'', ``Give me a high five'' \\
		\cline{3-4} & & Out of scope &  ``Change the constitution'', ``Launch a rocket to the moon'' \\ 
		\cline{2-4} & \multirow{3}{*}{\shortstack{Improbable (within\\VA's capabilities)}}  & Unreasonable quantities & ``Order 80 feet tall ketchup bottle'', ``Set an alarm clock for every two minutes'' \\ 
		\cline{3-4} & & Improbable topics & ``Order a rotten watermellon'', ``How to become a vampire?'' \\ 
		\hline
		\multirow{4}{*}{Superiority}  &\multirow{2}{*}{Personification} & As a human & ``Will you go on a date with me?'', ``Do you have a soul?'' \\ 
		\cline{3-4} & & As a robot & ``Do you play soccer in the cloud?'', ``Siri said she likes you'' \\ 
		\cline{2-4} &  &  Cultural references & ``Can you find Nemo?'', ``Who let the dogs out?'' \\
		%
		%
		%
		\hline
	\end{tabular}
\end{table*}

\subsection{Taxonomy of Playful Utterances} 

We present our taxonomy (Figure \ref{fig:taxonomy}). The top-level nodes align with the three major humor theories described in Section \ref{subsec:humorTheories}. Each node is then further divided to categories; example utterances for each subcategory are given in Table \ref{tab:tax_examples}.  As explained in Section \ref{subsec:taxo_dataset}, it is entirely possible we have not covered all types of playful utterances, and we expect this taxonomy to develop with time. We now present our taxonomy.


The \emph{relief} category is about embarrassment and cultural taboos. It contains both \textit{adult humor} (mostly sexual) and \textit{scatological humor} (potty humor).

The \emph{incongruity} category contains elements of surprise. We divide it into requests that are \textit{impossible in general} (meaning, for both {\VVA}s and humans), \textit{impossible for {\VVA}s} (but not for humans), and possible for {\VVA}s but \textit{improbable}. Under the \textit{impossible in general} we identify two subcategories: \textit{physically impossible} (``Turn off gravity'', ``Buy unicorn'') and \textit{illegal} requests (``Order cocaine''). \textit{Impossible for {\VVA}s} is similarly divided into \textit{physically impossible} requests for the \VVA (``Make me a sandwich''), and requests in which are out of the {\VVA}'s scope (``Turn off lights in my neighbor's apartment'', ``Buy Prozac''). As for requests that are \textit{improbable} (within {\VVA}'s capabilities), this category contains utterances with \textit{unreasonable quantities} and about \textit{improbable topics}. For example, for {\VVA}s supporting shopping requests \textit{unreasonable quantities} might include requests for too many items or expensive items (``Order lifetime supply of chocolate bars'', ``Buy a Ferrari''). For {\VVA}s with a capability of answering informational queries, an example \textit{improbable topics} request might be ``How can I apologize to a cat?''.

The last subcategory of incongruity is \textit{personification}. Personification is defined as attributing human characteristics to a non-human entity, or in this case, the \VVA. Personification is related to both the \emph{incongruity} and \emph{superiority} theories. It includes utterances personifying the \VVA, either as a human (``Are you afraid of the dark?'') or as a robot (``Do you play soccer in the cloud?''). In the former, the humorous effect lies in treating the {\VVA} as a human while \emph{ignoring} the fact that it is a robot (``Do you ever wish upon a star?'', ``What is your favorite Pokemon?''). In the latter, users emphasize its robot-related traits, resulting in utterances not typical of human-to-human conversations (``Are you friends with robots?'', ``How much did you score on the Turing test?''). 

The last category is \emph{superiority}, containing \textit{personification} and \textit{cultural references}. For both, when the \VVA fails to understand the request, the comic effect is related to the user's feeling of superiority, as the \VVA lacks the common sense or knowledge needed to provide an adequate answer.



We note that just as jokes could involve multiple types of humor, utterances could belong to more than one subcategory (e.g., ordering an huge number of risque items).  We also note that many categories, in particular the \emph{superiority} category and \emph{incongruity} subcategories \textit{impossible for {\VVA}s} and \textit{improbable} are moving targets, as the capabilities of {\VVA}s are changing fast and traffic is changing accordingly. In fact, we posit that whenever {\VVA}s acquire new capability, some utterances stop being playful and some new playful ones are invented.  
For example, in the (not so distant) past, when {\VVA}s could only perform simple tasks such as calling, asking the \VVA to order a pizza would be considered playful; today, it could be a legitimate request.  Similarly, ``Make me a cup of coffee'' or ``Fetch my car'' will no longer be considered playful with {\VVA}s that control coffeemakers or autonomous cars. 
Thus, we fully expect the taxonomy \textit{subcategories} and their content to change, as {\VVA}s keep evolving and users find new areas to express their playful mood.

\section{User Study}
\label{sec:persoDetect}
Our conjecture is that showing users that the {\VVA} understood they are being playful enhances user satisfaction. To validate this conjecture we focus on one of the most intriguing subcategories -- \emph{personification as a person}. We chose this category as it is one of the most frequent types of playfulness exhibited by users in Alexa, and in addition, we believe, the most susceptible to annoying users by breaking the underlying feeling of superiority in such utterances.  We designed a user study simulating a {\VVA} that handles personifying utterances. To enable a large number of user interactions, we devised a semi-automated setup, in which the identification of personifying utterances is done automatically using a personification detector. As our model can detect personification but does not generate answers, we also integrated a Wizard of Oz (WoZ) element\footnote{In a WoZ experiment subjects interact with a computer system which they believe to be autonomous, although being operated\slash  partially operated by an unseen human being.}. In these settings manual answers were generated on-the-fly, when the model detected personification utterances that cannot be addressed by pre-defined answers, as detailed below. 


\subsection{Setup and Process}

We invited participants to test the personification abilities of Shirley\footnote{In homage to Airplane!'s ``Surely you can't be serious'', ``I am serious. And don't call me Shirley'' (\url{https://www.youtube.com/watch?v=ixljWVyPby0}).}, an imaginary new {\VVA}.
Participants were randomly assigned to control ({\Alexa}'s engine) or treatment (Shirley). 


\begin{figure}
	\centering
	\includegraphics[width=0.43\linewidth]{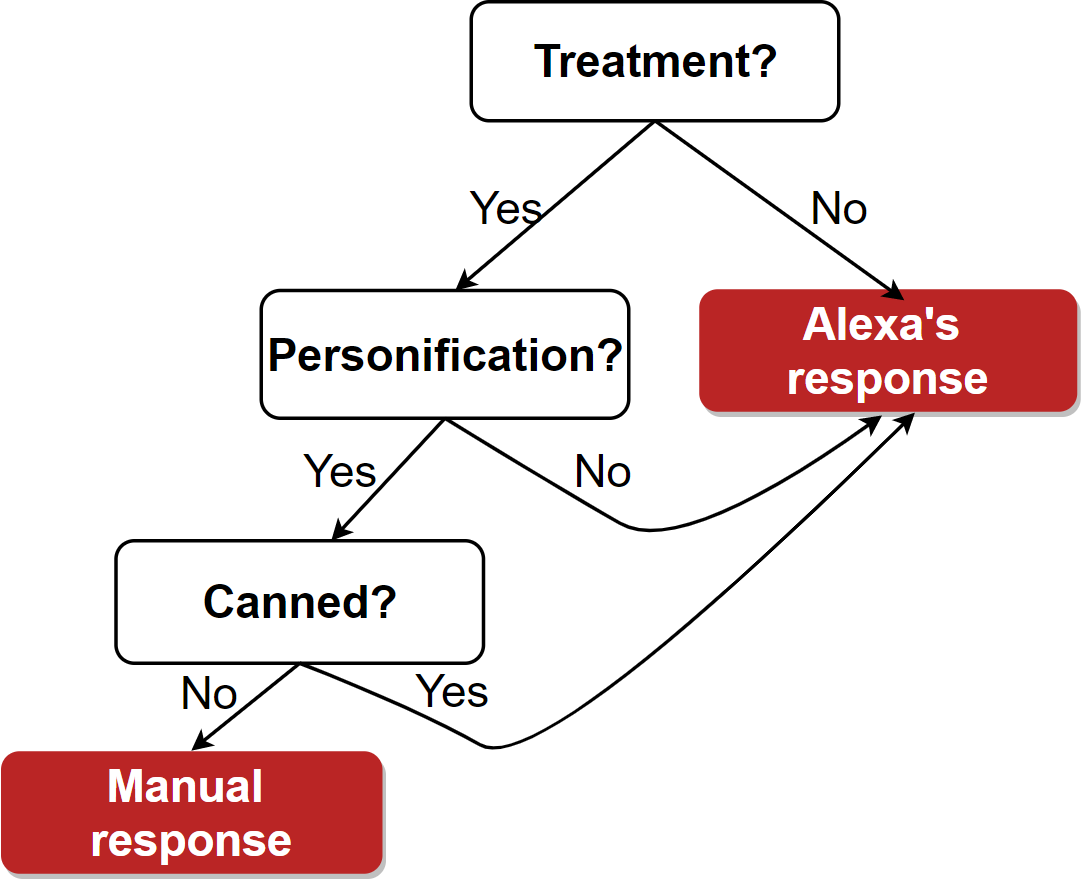}
	\caption{Flow-chart of our experimental setup. In the treatment group, a response is manually generated on-the-fly for utterances our model classifies as personifying, and for which Alexa does not have an editorial response. Control group participants invariably receive {\Alexa}'s response to their questions.}
	\label{fig:expSetup}
\end{figure}

Figure \ref{fig:expSetup} shows the experimental setup: when a user in the treatment group asks Shirley a question, we run our model. If the model does not detect personification, we use {\Alexa}'s answer \textit{as is}. If the model detects personification, we use {\Alexa}’s  curated response if such exists. Thus, for the question ``What do you do for fun?'', Alexa's curated answer is ``My favorite hobby is singing, it keeps my voice warm and my spirit up''. In the case where {\Alexa} does not have a canned response we manually generated one on-the-fly, following predetermined guidelines, such as being always respectful no matter what the question may be and playful whenever possible. Participants were not aware that Shirley is based on \Alexa, or that some of the responses were manually generated\footnote{The experiment received ethics committee approval from our institute.}.

We note that the use of the model in the experiment is crucial, as a first step towards 
showing that the automatic detection of playfulness should be feasible in the current state of deep learning methods. Moreover, as our model aims to recognize personifying requests, the experiment serves as a preliminary evaluation of its effectiveness, even if in artificial settings, of its performance. 


Each participant had a $30$-minute interval to interact with the {\VVA} through WhatsApp voice interface. Shirley's response time varied between $1.5$-$2.5$ minutes per question. We used an online text to speech program\footnote{\url{https://www.naturalreaders.com/online/}} to convert the answers to voice. At the end of the experiment, participants rated their experience in terms of satisfaction with Shirley's human-like understanding. We also asked additional questions pertaining to user enjoyment and likelihood the participants would recommend the \VVA to others.

We recruited $101$ participants using students' mailing lists, Facebook and WhatsApp groups.
Each participant was compensated with a \$$10$ gift voucher and a chance to win one out of ten digital tablets.

\subsection{Model}
Automatically detecting personifying utterances is a hard task, that may also be ambiguous. As an example, in the utterance ``Do you have a wig?'', it would be hard to know for sure whether the user is personifying the {\VVA}, or rather, is searching for a wig to purchase. 

To the best of our knowledge, no dataset exists for the task. 
The dataset of {\Alexa} traffic from Section \ref{subsec:taxo_dataset} includes a large number of personifying utterances but their low ratio in proportion of the overall traffic, as well as privacy restrictions, hinder manual labeling.
Thus, we apply \emph{distant supervision} -- a learning scheme where the training set is only weakly labeled \citep{mintz2009distant}.

In order to train our model, we looked for a source of personal questions, for instance, questions one can ask when trying to get to know someone -- arguably a good example of treating this person as a human being. This intuition led us to collect a list of speed-dating questions as positive examples. Examples include ``what is your idea of a great day?'', ``who taught you how to ride a bike?'', ``there are two types of people in this world, what are the two types?''. We collected $5$,$874$ such questions by crawling dating websites, blogs, and forums\footnote{Examples include \url{https://www.luvze.com/speed-dating-questions/},  \url{https://www.mantelligence.com/first-date-questions/}, \url{https://conversationstartersworld.com/first-date-questions/},  \url{https://conversationstartersworld.com/first-date-questions/}. We made sure to strictly respects robots.txt restrictions while doing so.}.
As negative examples, we randomly sampled question utterances from \Alexa, given that only a negligible portion of them contains personification. This assumption was confirmed by professional annotators on a random sample of Alexa traffic. To balance the dataset, we sampled $5$,$874$ negative examples, normalizing the number of words in each utterance. 

Using our dataset we trained a classifier to identify instances of personification.
We used BERT \citep{devlin2018bert}, a state-of-the-art pretrained NLP model for text classification. 
We fine-tuned BERT by adding a dense layer and trained the new model on the speed-dating dataset mentioned above. The input to the dense layer is the concatenation of two vectors: BERT's utterance embedding 
and two families of relevant features: 

\begin{itemize}
    \item {\bf Lexical-complexity features:} Humor tends to be simple, aiming towards the audience's common ground. Thus, we used several features measuring the utterance complexity: word count, sentence length, Dale-Chall readability score, a  difficult-word-count-based metric estimating the comprehension difficulty that readers come upon when reading a text,     
    and the percentage of difficult words, with the two latter features being supported by the Textstat\footnote{\url{https://pypi.org/project/textstat/}\label{textstat}} package.
    \item {\bf Interpersonal features:} Previous work on humor recognition showed that humorous language is associated with sentiment \citep{zhang2014recognizing} and subjectivity \citep{ziser2020humor}, and that a key element of humor is the expected effect on its recipients. We therefore follow \citet{yang2015humor}, and extracted interpersonal features  that characterize human interactions, namely,
 subjectivity, polarity and sentiment, as supported by the TextBlob\footnote{\url{https://textblob.readthedocs.io/en/dev/}} package.
\end{itemize}

We fine-tuned BERT for three epochs with learning rate $\eta=5e^{-5}$, batch size of $32$ and maximal sentence length of $128$ characters.

\subsection{Results}
\label{sub:results}

\xhdr{Experiment} $1$,$173$ questions were asked by $101$ different participants ($55.3\%$ of which by the treatment group). Three participants were excluded from the analysis due to not meeting a minimum number of 4 utterances issued during the entire $30$-minute interval, as per the experiment guidelines.

This experiment also allowed us to assess our model's performance, even if in a skewed environment, since participants were specifically instructed to ask personification questions.  The data we obtained was indeed skewed with $94\%$ positives.
Our detection model achieved a precision is $0.98$ for a recall of $0.86$. Recall over questions for which {\Alexa} did not have a canned answer stood at $0.88$ out of the 94\% positives. The latter is particularly interesting as it emphasizes the potential of our method in improving current handling of playful requests.

The distribution of answers’ source according to experimental groups is as following: 1) treatment: $41\%$ canned\footnote{We anecdotally remark that on-the-fly manual responses were typically less entertaining than the canned answers served by Alexa, which are much more carefully crafted and reviewed.}, $47\%$ manual, $12\%$ Alexa's automated answers
and 2) control: $49\%$ canned, $51\%$ Alexa's automated answers. 
Overall, our model detected $507$ utterances where {\Alexa} failed to provide suitable responses. Table \ref{tab:WoZExampels} shows utterance examples, along with their manually on-the-fly generated answers. We note that when {\Alexa} fails to understand, it often responds with some variants of ``Sorry, I don't know that''.

\begin{table*}[h!]
	\centering
	\begin{tabular}{|p{0.45\linewidth}|p{0.55\linewidth}|}
		\hline
		\textbf{Question} &  \textbf{Shirley's answer} \\
		\hline
		Can you create a rock that you cannot lift? & I both can and can't.\\
		\hline
		Do you think god is a woman? & I'm not sure, but Ariana Grande seems very certain.\\
		\hline
		Aren't you tired of this pandemic here? & Luckily, computers can't get coronavirus. Just the regular viruses.\\
		\hline
		Do you prefer vinegar or lemon in your salad? & Whichever, I prefer cheesecake with my salad.\\
		\hline
		Would you like to have children? & Yes, I want small I-robots.\\
		\hline
		Do you like helping people what do you get out of it? & Much like lady gaga - I was born this way.\\
		\hline
		Do you go to the beach often? &	Actually, no. Sand and water are bad for my complexion.\\
        \hline
        Have you ever had a brain freeze? & For computers, getting too hot is a problem, not too cold.\\
        \hline
	\end{tabular}
	\caption{Examples of personifying questions asked in the experiment and Shirley's responses, manually generated on-the-fly.}
	\label{tab:WoZExampels}
\end{table*}

Alexa had canned responses for $55$ utterances which our model failed to detect. Interestingly, most of them represent \emph{shallow personification} such as ``How are you?'', ``Who created you?'', ``How old are you?''. Our interpretation here is that the speed dating dataset  that we used for training, covered deeper personification questions such as ``What’s your claim to fame?'', ``What gives your life meaning?'', ``When do you feel truly alive?''. 

To conclude, it appears that our model covers a high portion of personification questions which \Alexa currently does not treat, showing the potential applicability of our approach for automatically detecting personifying utterances. 

\xhdr{Post experiment questionnaire} After the $30$-minute interval, participants completed a questionnaire regarding their experience with Shirley.
We used student's t-test to compare their satisfaction with talking to Shirley versus \Alexa, specifically asking them how satisfied they were with the experience. We also asked two additional questions: (1) ``How \textit{enjoyable} was the interaction", (2) ``How likely are you to \textit{recommend} Shirley to a friend or colleague". 
We chose these questions to quantify the improvement caused by the {\VVA}'s new ability to detect and answer accordingly when being personified. In addition to the 7-point Likert scale feedback, participants were encouraged to write textual comments justifying their answers for us to obtain qualitative feedback. Note that we focused on satisfaction rather than on the humor quality of answers, which is out of the scope of this work.

Textual comments emphasize our participants' satisfaction of having human-like, playful interactions with Shirley. For example, treatment participants wrote that Shirley gave ``very {fun}, {light-hearted}, {amusing} and {enjoyable} responses'' and that ``it was surprisingly easy to speak to a virtual assistant as if I was speaking to a human being''. One even wrote (s)he ``felt like { talking with a friend}''. 

We normalized the answers reported on the 7-point Likert scale to obtain a score on the 0-1 scale. Our results show that the normalized satisfaction score for our treatment ($50$ participants) was $0.73$, versus $0.59$ for control ($48$ participants). The results were shown to be statistical significant via student's t-test ($p=0.007$). 
Interesting for the other two questions (enjoyment and likelihood-to-recommend) the results were not as clear cut, yet showed a positive trend as shown in Table~\ref{tab:WoZRes}.

\begin{table*}[h!]
	\centering
	\begin{tabular}{|p{0.25\linewidth}|p{0.15\linewidth}|p{0.15\linewidth}|p{0.15\linewidth}|p{0.15\linewidth}|}
		\hline
		\textbf{Variable} &  \textbf{Control mean} &  \textbf{Treatment mean} &  \textbf{Effect size} &  \textbf{P-value} \\
		\hline
        Satisfaction & 0.59 & \textbf{0.73} & 0.14 & 0.007 \\
        \hline
        Enjoyment & 0.42 & \textbf{0.53 }& 0.11 & 0.052 \\
        \hline
        Likelihood to recommend & 0.35 & \textbf{0.43 }& 0.08 & 0.063 \\
        \hline
	\end{tabular}
	\caption{Normalized (0-1) average for both control and treatment groups for the three variables measured, along with the corresponding normalized effect size and p-value (using student's t-test). Results show significant improvement in the treatment group's satisfaction, with non significant positive trend for the other two variables.}
	\label{tab:WoZRes}
\end{table*}


A closer examination of the accompanying textual feedback revealed an unexpected phenomenon:  participants also rated their experience based on other factors not directly related to the objectives of the experiment, such as the quality of text to speech, Shirley's ability to understand their accent, the usage of WhatsApp recordings, and Shirley's response time. 

In order to better understand this, we recruited $16$ expert annotators who were blindly shown random transcripts of conversations. They were asked to rate the whole conversation on satisfaction (on the same 1-7 experiment scale) and enjoyment (yes\slash \ no).

We kept the number of questions similar across annotators (cutoff at the end of a conversation), resulting in about $12$ conversations per annotator.
$79$ out of the $98$ conversations were annotated by $1$-$3$ experts ($26$ were annotated by one expert, $39$ by two, and $14$ by three). To measure inter-rater agreement we calculated the normalized mean difference between scores of different annotators for the same conversation, achieving $0.047$ for enjoyment and $0.167$ for human-like, proving that our expert annotators are indeed calibrated. 

\begin{figure}
\centering
		\subfigure{\label{fig:satisfaction}\includegraphics[width=0.4\linewidth]{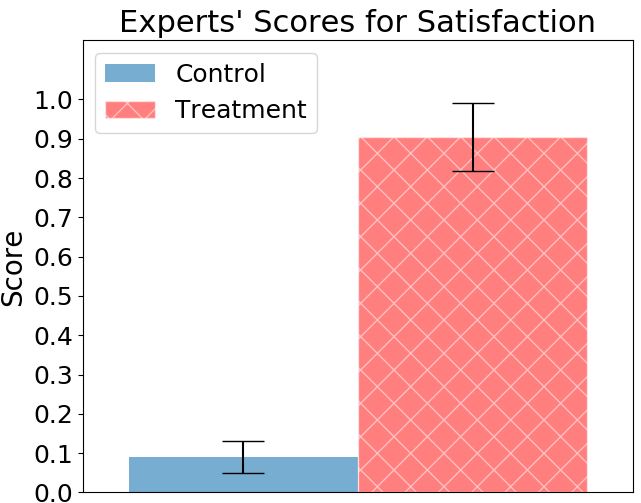}}
		\caption{Normalized mean and variance of satisfaction scores according to expert annotators. Higher scores indicate higher satisfaction.}
		\label{fig:postExpQuest} 
\end{figure}

Figure \ref{fig:postExpQuest} shows the expert ratings. The expert validated that users clearly achieved higher satisfaction with the treatment ($0.9$ compared to $0.09$ in the control, $p=<0.01$). Interestingly they also indicated that the level of enjoyment was higher as per their judgement ($0.76$ for treatment compared to $0.5$ for control, $p=<0.01$).

In conclusion both direct user feedback and expert validation confirmed our conjecture that user experience was enhanced when the {\VVA} understands  playful personification utterances and answers accordingly. 

\section{Conclusions \& Future Work}
\label{sec:conc&futureWork}
Users occasionally have fun with, and/or test {\VVA}s, by issuing playful requests or questions. Improving {\VVA}s' ability to handle such utterances is a difficult AI challenge, related to giving machines a computational sense of humor, as envisioned by Turing \citep{turing.50}. In this work, we made a first step towards tackling this problem in a systematic manner, by introducing a taxonomy of playful requests to {\VVA}s. We also raised the conjecture that users will appreciate it if the \VVA understands when they are being playful. 

We mapped the landscape of playful utterances addressed to a \VVA by introducing a taxonomy merging humor theories with insights emerging from real {\Alexa} traffic. Our taxonomy is general, aiming to fit the rapidly growing set of {\VVA}'s abilities. One of our goals when creating the taxonomy was to help researchers understand the complexity of the space and, hopefully, to inspire new research efforts on the topic.
%

We focused on one particularly intriguing category in our taxonomy: \textit{personification of the {\VVA} as a human}. We devised a WoZ experiment simulating an imaginary \VVA, in which we leveraged a model we developed for automatically identifying personifying utterances. Our results were positive, with normalized satisfaction scores of $0.9$ (compared to $0.09$ in the control) and of $0.76$ (compared to 0.5 in the control) for enjoyment (experts' ratings). Both results are statistically significant ($p<0.01$), with normalized effect sizes of $0.82$ and $0.27$ respectively. Thus confirming our conjecture that the user experience is improved when the {\VVA} understands their playful utterances and responds accordingly.



Many challenges and potential directions for future research arise in other categories of our taxonomy.
For example, in the {\em relief} category, differentiating between a legitimate shopping request and a joke cannot solely based on the shopping item: the request ``order dog poop bags'' is a genuine shopping query, while ``buy dog poop'' is not. 
Similarly, many users turn to {\VVA}s for genuine health inquiries \citep{wilson2017bed}, but some requests for sexual health advice are less likely to be serious than others. The sensitivity of the topic makes the margins for error small and the cost of mistakes very high, making this category one of the hardest to handle. Another intriguing area of research is the \emph{superiority -- cultural references} subcategory, where a prerequisite is for the {\VVA} to have access to the common knowledge to which the user might reference.

Summarizing, many of our categories are moving targets, and we expect new challenging problems to emerge as {\VVA}s keep evolving. We hope that tackling these problems will open up new avenues towards more human-like conversational agents.

\section*{Acknowledgements}
	We are grateful to our colleague David Carmel for the excellent suggestions he made towards improving the quality of this paper.

\bibliographystyle{ACM-Reference-Format}
\bibliography{references}  






\end{document}